\documentclass[12pt,titlepage]{article}
\usepackage[pdftex,bookmarks,pagebackref, colorlinks = true, 
linkcolor=blue, urlcolor = blue,
pdfpagemode=UseOutlines]{hyperref}
\usepackage[T1]{fontenc}
\usepackage[latin1]{inputenc}
\usepackage[dvips]{graphicx}
\usepackage[english]{babel}
\usepackage{harvard,ae,amsmath, amsfonts, lscape,mathabx}
\usepackage{ae,amsmath, amsfonts, lscape,mathabx}
\usepackage{bm,booktabs}
\usepackage{caption}
\usepackage{appendix}
\usepackage{pdfpages}
\usepackage{tikz}
\usepackage{blkarray}
\usepackage{geometry}
\usepackage{multirow}
\usepackage{subfloat}

\newcommand{\can}[1]{\citeasnoun{#1}}

\usetikzlibrary{shapes, arrows, calc, positioning}
\everymath{\displaystyle}
\definecolor{orange}{RGB}{255,127,0}
\definecolor{cmt}{RGB}{85,170,170}

\newcommand{\commentbs}[1]{}

\newcommand{\bslpp}[1]{\begin{slide}[Wipe]{{\small #1}}}
\newcommand{\eslpp}{\end{slide}}
\newcommand{\bitst}{\begin{itemstep}
         \item[]
         \vspace{-4mm}
}
\newcommand{\eitst}{\end{itemstep}}
\def\tfrac#1/#2{\leavevmode
        \kern.1em \raise .5ex \hbox{\the\scriptfont0 #1}%
        \kern-.1em $/$%
        \kern-.15em \lower .25ex \hbox{\the\scriptfont0 #2}}
\def\makeactive#1{\catcode`#1=\active \ignorespaces}
\def\alwaysspace{\hglue\fontdimen2\the\font \relax}%
{\makeactive\^^M \makeactive\ %
\gdef\obeywhitespace{%
\makeactive\^^M\def^^M{\par\indent}%
\makeactive\ \let =\alwaysspace}}%
\newcommand{\dpart}[2]{\frac{\partial #1}{\partial #2}}

\def\bkR{{\rm I\kern-.17em R}}
\def\bkQ{{\rm I\kern-.17em Q}}
\def\bkN{{\rm I\kern-.17em N}}
\def\bkZ{{\rm I\kern-.17em Z}}

\def\uniset{{\rm 1\kern-.40em 1}}
\newenvironment{stdarray}{\[  \left\{ \begin{array}{lcl}}{\end{array} \right. \]}
\newcommand{\bstd}{\begin{stdarray}}
\newcommand{\estd}{\end{stdarray}}
\newcommand{\beq}{\begin{equation}}
\newcommand{\eeq}{\end{equation}}
\newcommand{\erps}{\varepsilon}

\newcommand{\bc}{\begin{center}}
\newcommand{\ec}{\end{center}}
\graphicspath{{./}}

\newcommand{\incgraf}[3]{
\begin{figure}
\begin{center}
\includegraphics[#3]{#1}
\end{center}
\caption{#2} \label{#1}
\end{figure}
}

\newcommand{\incgraff}[4]{
\begin{figure}
\begin{center}
\includegraphics[#4]{#1}

\vspace{0.25cm}

\includegraphics[#4]{#2}
\end{center}
\caption{#3} \label{#1}
\end{figure}
}

\newcommand{\incgraffapp}[4]{
\begin{figure}[h!]
\begin{center}
\includegraphics[#4]{#1}

\vspace{0.25cm}

\includegraphics[#4]{#2}
\end{center}
\vspace{-0.5cm}
\caption{#3} \label{#1}
\end{figure}
}

\begin{document}

\title{\Large\textbf{The Econometrics of Matching with Transferable Utility:\\ A Progress Report\thanks{We thank Alfred Galichon, Jeremy Fox, Thierry Magnac, and Shruti Sinha for helpful comments. A previous version of this paper was circulated under the title ``Matching with Random Components: Simulations''.  We dedicate this paper to the memory of our friend Yinghua He.}}}

\author{Pierre-Andr\'e Chiappori\footnote{Department of Economics, Columbia University. Email: \href{mailto:pc2167@columbia.edu}{pc2167@columbia.edu}}
\and Dam Linh Nguyen\footnote{Department of Economics, New York University. Email: \href{mailto:n.linh@nyu.edu}{n.linh@nyu.edu}} \and 
Bernard Salani\'e\footnote{Department of Economics, Columbia University. Email: \href{mailto:bsalanie@columbia.edu}{bsalanie@columbia.edu}}
}

\date{\today}

\maketitle

\begin{abstract}
Since \can{Choo-Siow:06}, a burgeoning literature  has analyzed matching markets when utility is perfectly
 transferable and
 the  joint surplus is separable. We take stock of recent methodological
 developments in this area. Combining theoretical arguments and simulations, we show that the separable approach is reasonably robust to omitted variables and/or non-separabilities.  We conclude with a caveat on data requirements and imbalanced datasets.
\end{abstract}

\section*{Introduction}
The empirical analysis of matching markets has made considerable progress in recent years.
 We will focus 
here on 
markets where partners exchange transfers freely. In the usual
 terminology,
 we study matching markets
 with perfectly transferable utility (hereafter ``TU''). 
In most matching markets, observationally identical agents end up with very different matching outcomes: some may be 
unmatched, and others will be matched to partners with variable observable characteristics. This can be
 rationalized by introducing frictions, unobserved heterogeneity, or a mixture of both. 
As explained in \can{PABmatchingsurvey}, these two groups of rationalizations have essentially identical predictions for
cross-sectional data; they can only be distinguished with data featuring transitions, which are often 
unavailable to the analyst. We choose here to model the dispersion of outcomes using unobserved heterogeneity. 
More precisely, we work with the class of models pioneered by \can{Choo-Siow:06}, which incorporates a 
quasi-additive error structure called {\em separability\/}
 by \can{csw:17}.  \citename{galsalpp17}~\citeyear{galsalpp17,cupid:22} 
study the properties of separable models in much more detail. \can{estsep:24} 
show that this class of models can be easily estimated using minimum distance, which often
 boils
down to generalized least squares; and that the Choo and Siow model can also be estimated with Poisson GLM.

 Beyond \can{Choo-Siow:06}, this framework has 
been applied by \can{csw:17} to changes in the marital college 
premium in the US. \can{soyoon:jole} used it to analyze cross-border marriages in South-East Asia, and \can{assortmatchincome} to describe matching on income levels in the marriage market.  It was also applied to the (many-to-many) market for car parts by \can{fox:qe18};  to  the labor market (in a many-to-one version) by \can{corblet:22}; and to mergers and acquisitions (a roommate problem) by \can{grst:24}.   It was extended to continuous types  by \can{dg:14} 
to study the contribution of personality traits to marital
 surplus;
 by \can{gkw} to imperfectly transferable utilities;  and by 
 \can{cgg:19} to non-bipartite matching in order to 
 compare same-sex and different-sex marriages.  

 By definition, separability  rules out the interaction between partners' unobserved heterogeneity 
in the production of joint surplus. While it is a very useful assumption for keeping the model tractable, it is also
 restrictive, especially
if the set of  characteristics available to the analyst is small. A standard intuition would suggest that this may
not matter much if unobserved characteristics are drawn independently of observed 
characteristics. However, we 
 are dealing here with a two-sided market where we cannot just transpose
this  intuition. 
The first   goal of this paper is to explore
 the consequences of relaxing separability on equilibrium matching
patterns, utilities, and division of surplus. We will also quantify the misspecifications that result from
mistakenly assuming separability when estimating a non-separable model.

Theoretical arguments and a Monte Carlo simulation suggest  that  non-separability
impacts matching patterns and utilities in ways that, ex post, seem intuitive. It takes
a rather large  amount of non-separability to see a qualitative difference from the separable case, however. 
We find that when we estimate a non-separable model as if it were separable, 
the estimated complementarities in surplus are surprisingly robust. This is reassuring since 
we have known since \can{becker:73} that complementarities play a crucial role in TU matching markets.
Our conclusions are strongest when the distributions of the observable characteristics of the partners are 
similar (``symmetric margins''). As we will see, non-separabilities matter more when  markets
 are very unbalanced.

The second contribution of the paper is to point out that estimates may be fragile when the data contain matching cells of very different size. We give a simple, two-by-two example to show that the asymptotic variance of the estimator may be dominated by the inverse of the size of the smallest cell, unless precautions are taken to minimize its influence.
 
The
remainder of the paper is organized as follows. Section~\ref{sec:set_up}
 introduces the class of separable matching models with perfectly transferable utilities. Section~\ref{sec:discussion_of_separability} discusses the separability assumption and its implications.
Section~\ref{sec:mispecifying_separability}
explores the consequences of omitted variables and of mistakenly imposing separability.  Section~\ref{sec:caveat_on_imbalanced_data}
discusses the effects of imbalanced data, and how to remedy them.
We conclude with some directions for further work.

\section{Model and Methods}\label{sec:set_up}
We focus throughout on a bipartite one-to-one model of matching with unobserved heterogeneity and
perfectly transferable utilities\footnote{\can{estsep:24} extend the analysis to more complex separable models.}. Since it is similar to Becker's ``marriage market'', we will use that terminology and refer
to potential partners as  ``men'' and ``women''. Our  framework easily accommodates other interpretations.
 We maintain some of the standard assumptions: 
  matching is frictionless and all potential partners have the same information. The analyst only observes 
  a subset of  individual characteristics. 
  
 Men are indexed by $i\in\mathcal{I}$ and women by $j\in\mathcal{J}$. 
 Each man $i$ (resp.\ woman $j$) has a ``type'' $x_i\in \mathcal{X}$ (resp.\ $y_j\in\mathcal{Y}$). This is observable to all men,
 all women, and to the analyst. 
  When a man $i$ and a
woman $j$ marry, their match generates a   surplus 
$\tilde{\Phi}_{ij}$ which they can share freely.
 Singles match to ``0'': a single man
$i$ attains a utility of $\tilde{\Phi}_{i0}$ and a single
woman $j$ attains $\tilde{\Phi}_{0j}$. 

This paper will assume that the sets of types are finite: $\mathcal{X}=\{1,\ldots,X\}$ and $\mathcal{Y}=\{1,\ldots,Y\}$.  As mentioned in the introduction, \can{dg:14} extend the model to continuous types.
 We will denote by $\bm{\tilde{\Phi}}=(\tilde{\Phi}_{ij})_{i=1,\ldots,I; j=1,\ldots,J}$ the matrix of surpluses. 

 A matching is a list of numbers $(\mu_{ij}, \mu_{i0}, \mu_{0j})_{i=1,\ldots,I; j=1,\ldots,J}$, with $\mu_{ij}=1$ if $i$ and $j$ match, $\mu_{i0}=1$ if $i$ is single, etc.
We focus on the stable matching (it is generically unique). Since only types are observed by the analyst, the data consist of  {\em matching patterns}: the numbers of men and women of each type, which we denote $n_x$ and $m_y$, and the numbers of marriages $\mu_{xy}$ between  men of each type $x$ and women of each type $y$. By construction, $\textstyle\mu_{x0}=n_x-\sum_{y=1}^Y \mu_{xy}$ and $\textstyle\mu_{0y}=m_y-\sum_{x=1}^X \mu_{xy}$ are the numbers of single men of type $x$ and of single women of type $y$. Note that in some empirical applications, the analyst may not observe singles---only the numbers of marriages $\mu_{xy}$.

We denote by $\bm{\Phi}$ and $\bm{\mu}$ the matrix of surpluses and the matching patterns, respectively; and $\bm{r}=(\bm{n},\bm{m})$ the {\em margins}, i.e., the numbers of men and women of each type. Like most of the literature, we work with the {\em large market limit\/} in which the vectors $\bm{n}$ and $\bm{m}$ grow proportionately.

\subsection{Separability}\label{sub:separability}
The separable model restricts the form of the surplus $\tilde{\Phi}_{ij}.$ It imposes that there exist a matrix
$\bm{\Phi} =(\Phi_{xy})_{x\in\mathcal{X},y\in\mathcal{Y}}$, random variables $\bm{\varepsilon^i}=(\varepsilon^i_0,\varepsilon^i_1, \ldots,\varepsilon^i_Y)$, and 
$\bm{\eta^j}=(\eta_0^j, \eta_1^j, \ldots, \eta^j_X)$ such that
\begin{equation}
	\label{eq:sep}
	\tilde{\Phi}_{ij}  \coloneq 
			\Phi_{x_{i},y_{j}} 
				+  \varepsilon^{i}_{y_j}+ \eta^{j}_{x_i}
\end{equation}
within any match, and that single men and women  get $\tilde{\Phi}_{i0}=\varepsilon^i_0$ and
 $\tilde{\Phi}_{0j}=\eta^j_0$, respectively.
Moreover, the random vectors $\bm{\varepsilon^i}$ (resp.\ $\bm{\eta^j}$) are assumed to be drawn independently 
of each other, conditional on $x_i$ (resp.\ $y_j$).

\can{csw:17} and \citename{galsalpp17}~\citeyear{galsalpp17,cupid:22,estsep:24} 
studied the class of separable models in  depth. They derived some of their properties, proved
identification results, and proposed estimators.   We summarize them briefly here and refer the reader to recent surveys by \can{bs:hdbkem1} and \can{agbs:hdbkec} for more information.

\subsubsection{Identification}
We assume in this paper that the available data only describe ``who matches with whom'': that is, they consist of the matching patterns $\bm{\mu}$. Data on transfers between partners and/or proxies for the outcomes of matches, if available, may   provide more identifying power\footnote{See \can{salanieidentobstransfers:15}  for the case when transfers are observed.}.

\can{cupid:22} show that the stable matching patterns maximize the social welfare (the sum of the surpluses of all matches) 
\[
\sum_{x=1}^X\sum_{y=1}^Y \mu_{xy}\Phi_{xy} +\mathcal{E}(\bm{\mu})
\]
where the {\em generalized entropy\/} $\mathcal{E}$  depends on the margins $\bm{r}$ and on the distributions of the $\varepsilon^i$ and $\eta^j$ terms. This is a globally strictly convex problem. When (as we will assume) these terms have full support,  all matching patterns must be strictly positive and the set of first-order conditions
\begin{equation}\label{eq:identphi}
  \Phi_{xy}=-\dpart{\mathcal{E}}{\mu_{xy}}(\bm{\mu})
\end{equation}
defines what can be identified from a single matching market.

For any choice of the distributions of the $\varepsilon^i$ and $\eta^j$ terms, the joint surplus matrix $\bm{\Phi}$ is just identified. If for instance the analyst wants to identify the standard errors  of these terms, restrictions on the joint surplus must be imposed. The alternative is to pool data from several  markets and impose cross-market restrictions. This was the approach of \can{csw:17}, for instance.  

When the analyst only observes realized matches (no data on singles), a similar set of equations apply; but they only identify the joint surplus matrix up to arbitrary additive transformations $\Phi_{xy}\to \Phi_{xy}+a_x+b_y$. This simply translates the fact that the value of marriage cannot be identified without data on the proportion of singles.

\subsubsection{Inference}
Now suppose that the distributions of the unobserved heterogeneity terms (resp.\ the joint surplus matrix $\bm{\Phi}$) are known up to a parameter vector $\bm{\alpha}$  (resp.\ $\bm{\beta}$), and that the model is identified from data  on a single market\footnote{This is typically the case if the dimensions of $\bm{\alpha}$ and $\bm{\beta}$ add up to $X\times Y$ or less.}. For any value of $\bm{\alpha}$, the generalized entropy $\mathcal{E}^{\bm{\alpha}}$ can be evaluated easily---often in closed form. Given observed matching patterns $\bm{\hat{\mu}}$ ,  the parameters $(\bm{\alpha},\bm{\beta})$ can be estimated by minimizing the norm of the $X\times Y$ vector with components
\[
\Phi^{\bm{\beta}}_{xy} +\dpart{\mathcal{E}^{\bm{\alpha}}}{\mu_{xy}}(\bm{\hat{\mu}}).
\]
The resulting  minimum distance estimator  has the usual properties: it is consistent and asymptotically normal; there is  a  choice of the norm that minimizes its  variance-covariance matrix; and for that choice of the norm, the minimized value of the
objective function provides a $\chi^2$ test of the specification. 

While we focus here on parametric models, we note that \can{gualdanisinha:jpe23} have explored nonparametric inference in this context. \can{fox:qe18} has also  proposed a maximum-score estimator that he applied to a many-to-many Choo and Siow market.

\subsection{The Choo and Siow Specification}\label{sub:the_choo_and_siow_specification}

The best-known and most convenient separable model has
 the ``multinomial logit'' form popularized by \can{Choo-Siow:06}.   They assumed that each component of the vectors 
 $\bm{\varepsilon^i}$ and  $\bm{\eta^j}$
  is drawn independently from a centered standard type I extreme value distribution.
   They showed that the stable matching patterns  $\bm{\mu}$ obey very simple formul\ae: 
   the number $\mu_{xy}$ of marriages between men  of education $x$ and  women of education $y$ can be written as
   \[
   \mu_{xy} = \sqrt{\mu_{x0}\mu_{0y}}\exp(\Phi_{xy}/2)
   \]
  where  $\mu_{x0}$ (resp.\ $\mu_{0y}$) is the number of men of education $x$ 
  (resp.\ women of education $y$) who are single in equilibrium.
  
    This gives $X\times Y$ equations in $X\times Y+X+Y$ unknowns; the system is completed by the scarcity constraints
    \begin{align*}
       n_x &=  \sum_{y=1}^Y \mu_{xy} +\mu_{x0}  \\
  	      m_y &=  \sum_{x=1}^X \mu_{xy} +\mu_{0y}.
    \end{align*}
  \can{cupid:22} show how  given the ``margins'' $(n_x), (m_y)$ and the joint surplus matrix $\bm{\Phi}$, 
    this system of equations can be solved very efficiently using an I\-te\-rative Projection Fitting Procedure (IPFP). 
    Identification is straightforward: since the distributions of the unobserved heterogeneity terms are parameter-free, the joint surplus matrix $\bm{\Phi}$ is just identified from the matching patterns $\bm{\mu}$. In fact, the ge\-neralized entropy in this case is the usual entropy, and the first-order conditions~\eqref{eq:identphi} give
    \[
    \Phi_{xy} = \log\frac{\mu_{xy}^2}{\mu_{x0}\mu_{0y}}.
    \]
    When the numbers of singles are not observed, $\Phi_{xy}=2\log\mu_{xy}+a_x+b_y$, with arbitrary $\bm{a}$ and $\bm{b}$.
In addition to the minimum distance procedure described earlier, a Poisson model with two-way fixed effects can be used to estimate the Choo and Siow model \citeaffixed{estsep:24}{see}.

  Since the \can{Choo-Siow:06} specification is so simple and has been widely used, 
  it is a natural benchmark and we will use it in our examples.
  It  has been criticized on a 
number of grounds, however; see for instance 
\citename{galsalpp17}~\citeyear{galsalpp17,galsalmatchingiia:19,cupid:22}. Since the components of $\bm{\varepsilon^i}$ are independent of each other, the specification rules out ``local'' correlations. In  addition, the Choo and Siow matching model inherits a two-sided version of the Independence of Irrelevant Alternatives property of the one-sided multinomial logit. As in the one-sided case, this has some advantages and many drawbacks\footnote{See \can{mourifieet19} and \can{mourifiesiow}
  for extensions of the Choo and Siow framework that reach beyond separable models.}.

\section{Discussion of Separability}\label{sec:discussion_of_separability}
By definition, separability rules out any contribution to the joint surplus of a match 
of interactions between 
characteristics of partners that are unobserved by the analyst.  
Separability is best viewed through the lens of an ANOVA decomposition: it requires that conditional on the observed types of the partners, interactions between their unobserved types do not contribute to the variation in the surplus. 

Separability does {\em not\/} rule out ``matching on unobservables'': a man $i$ of education $x$ and a woman $j$ of type  $y$ are more likely to marry if his $\erps^i_y$ and her $\eta^j_x$ take  higher values. It does have strong consequences, however.  Suppose $i$ and $j$  do marry at the stable matching. Since this man  gets utility $U_{xy}+\erps^i_y$, he 
would be equally happy with any other woman of type $y$. If for instance the observed types only record education, and 1,000 couples form between college-educated partners, one could shuffle these 1,000 men and 1,000 women without changing anyone's utility level.  

As can be seen from this example, separability
   is more restrictive when the data contain little information on types and/or  unobserved heterogeneity  is highly relevant for the application under study. In a sense, assuming separability acknowledges that in these two-sided problems, the data are unlikely to give us much information about  interactions between unobserved components\footnote{But see \can{foxyanghsu} for some results along these lines.}. Still, it is worth investigating whether separable models yield robust conclusions.

Separability has very useful properties that greatly simplify  analysis and inference. It implies that 
at the stable matching, there exist matrices $\bm{U}$ and $\bm{V}=\bm{\Phi}-\bm{U}$ such that if a man $i$ of type  $x$ and a woman $j$ of type $y$  marry in equilibrium, he gets utility
\[
u_i=U_{xy}+\erps^i_y=\max\left(\erps^i_0, \max_{t=1,\ldots,Y} (U_{xt}+\erps^i_t)\right)
\]
and she gets $v_j=V_{xy}+\eta^j_x=\max\left(\eta^j_0, \max_{z=1,\ldots,Y} (V_{zy}+\eta^j_z)\right)$. 

The stable matching $\bm{\mu}$ is homogeneous of degree 1 in the vector $(\bm{n},\bm{m})$:  if the numbers of men and women of each type are multiplied by a common positive integer $k$, then all numbers of marriages  $\mu_{xy}$ and  of singles $\mu_{x0}$ and $\mu_{0y}$ are multiplied by $k$ as well. On the other hand, utilities are unchanged; in that sense, there is no scale effect in a separable matching market.

As the matching patterns maximize  the social welfare, which is globally convex, its Hessian must be semi-definite positive at the optimum. This entails testable implications that (depending on the available data) may allow for tests of the separability assumption \citeaffixed{cupid:22}{see}.

\section{Misspecifying Separability}\label{sec:mispecifying_separability}
We  explore here the consequences of mistakenly imposing
separability. There are of course many ways to add
 non-separable components to a separable matching model.  We could for instance add a simple interaction term to Equation \eqref{eq:sep}:
 \begin{equation}\label{eq:nonsep_interaction}
	\tilde{\Phi}_{ij}  \coloneq 
			\Phi_{x_{i},y_{j}} 
				+  \tau(\varepsilon^{i}_{y_j}+ \eta^{j}_{x_i})+\sigma  \; \bm{\xi}_i\cdot\bm{\zeta}_j
\end{equation}
with $\tau$ and $\sigma$ two positive parameters that control the relative strength of the non-separable component. We call this a case of ``missing interaction''.

Alternatively, we could  use a ``missing shock'' model. Generate a  matrix 
 $\bm{\nu}=(\nu_{ij})_{i=1,\ldots,I; j=1,\ldots, J}$  of independent draws from some mean-zero 
 distribution and  take $\nu_{ij}$ to represent a pair-specific unobserved preference 
 shock:
 \begin{equation}\label{eq:nonsep}
	\tilde{\Phi}_{ij}  \coloneq 
			\Phi_{x_{i},y_{j}} 
				+  \tau(\varepsilon^{i}_{y_j}+ \eta^{j}_{x_i})+\sigma    \nu_{ij}.
 \end{equation}
 In both cases,  
we recover the separable surplus in \eqref{eq:sep} if we set 
 $\tau=1$ and $\sigma=0$.

 \subsection{Missing Interactions}
Missing interactions may not create major difficulties, as long as they are only weakly correlated  to the included interactions.  
Suppose that  the true model is as in Choo and Siow, with one difference:  the types $\bm{x}$ and $\bm{y}$ are continuous variables  distributed as  two zero-mean Gaussian random variables in $\mathbb{R}^{K+1}$.  We take the joint surplus to have the following quadratic form:
$$
\Phi(x,y)=\sum_{k, l=1}^{K+1} A_{kl} x_k y_l.
$$
This is a version  of the quadratic-normal \can{tinbergen:56} model. 
It follows from  \can{bojilovgalichon:16} that  the stable matching is described by an affine mapping between men and women types:  $\bm{y}=\bm{T}\bm{x}+\bm{\xi}$  where conditional on $\bm{x}$, $\bm{\xi}$ has mean zero. The matrix  $\bm{T}$ depends on the matrix $\bm{A}$ and the variance-covariance matrices $\bm{\Sigma_x}$ and $\bm{\Sigma_y}$ of $\bm{x}$ and $\bm{y}$. The matrix $\bm{T}$ can be estimated simply by regressing $\bm{y}$ on $\bm{x}$ over observed matches. Moreover, given $\bm{\Sigma_x}$ and $\bm{\Sigma_y}$ (which are easily estimated), the mapping from $\bm{A}$  to $\bm{T}$ can be inverted to recover an estimator of $\bm{a}$. 

Now suppose that   there are no interactions between $x_{K+1}$ and $(y_1,\ldots,y_K)$, as well as between $y_{K+1}$ and $(x_1,\ldots,x_K)$, so that the matrix $\bm{A}$ has the same block-diagonal structure:
\[
\Phi(x,y)=\sum_{k, l=1}^{K} A_{kl} x_k y_l + A_{K+1,K+1} x_{K+1}y_{K+1}.
\]
If we only observe $(x_1,\ldots,x_K)$ and $(y_1,\ldots,y_K)$, can we still estimate consistently the coefficients $(A_{11},\ldots,A_{KK})$?  This obviously won't work in general: if for instance $E(x_{K+1}\vert x_1,\ldots,x_K)=x_1$ and $E(y_{K+1}\vert y_1,\ldots,y_K)=y_1$, then omitting $x_{K+1}$ and $y_{K+1}$ will bias the  estimate of the coefficient $A_{11}$.  However, if we further assume that the $(K+1)$-th components of $\bm{x}$ and $\bm{y}$ are independent of the first $K$ components, the answer is positive\footnote{If some omitted variables are {\em not\/} independent of the included variables, one could conceivably use instrumental variables, as is done in a linear model. We leave this for further research.}. This is easy  to see from the formul\ae\ in \citeasnoun[Theorem 1]{bojilovgalichon:16}: under our assumptions, the  matrix $\bm{T}$ inherits the  block-diagonal structure of $\bm{\Sigma}$,  and   its top-left $(K,K)$ block does not depend on the value of $A_{K+1,K+1}$ or of the variances of $x_{K+1}$ and $y_{K+1}$. As a consequence, the coefficients $(A_{11},\ldots,A_{KK})$ can be estimated by regressing the vector $(y_1,\ldots,y_K)$ on the vector $(x_1,\ldots,x_K)$.

It follows from this result that if we do not observe $x_{K+1}$ or $y_{K+1}$, we can still get consistent estimates of $(A_{11},\ldots,A_{KK})$. 
This finding obviously extends to more general block structures. It is superficially similar to  classic results on (the absence of) omitted variable bias in the linear model: it is much less general, however. It only obtains here  because  with a quadratic joint surplus and Gaussian types, the ``optimal transport map'' $\bm{y}=\bm{T}(\bm{x})$ yields a linear statistical model. Still, this suggests that the separable model has some degree of robustness to omitted variables that are functionally and statistically independent of those that are included.

\subsection{Missing Shocks}
Let us now turn to the model of \eqref{eq:nonsep}, where the true joint surplus includes pair-specific unobserved preference 
 shocks that are excluded from the estimated model.  We explored this case extensively via Monte Carlo simulations. 
We assumed
that the $\varepsilon$ and $\eta$ terms are drawn independently from a standard type I extreme value distribution, as in the \can{Choo-Siow:06} model.
 As is well-known, the surplus matrix $\bm{\Phi}$ is only identified up to a scale factor in the Choo and Siow model. This 
 factor equals the  standard error of the sum of the $\varepsilon$ and $\eta$ terms.
The total variance of the surplus in the separable ($\sigma=0, \tau=1$) model is  $\pi^2/3$ (twice the variance of a standard type-I EV term);  we ensure
  that it is the same in the non-separable model by imposing $\tau^2=1-\sigma^2/2$. This implies that 
$\sigma$ must take values in $[0, \sqrt{2}]$. To put it differently, it is tempting to define a ``coefficient of determination'' by 
$R^{2}=\sigma^2/2$. It represents 
the fraction of the  total variation in the joint  surplus that 
is explained by the non-separable component
 (the pair-specific individual preferences $\nu_{ij}$)  for 
 given $x_i$ and $y_j$.
 Our simulations will cover the whole range from $R^2=0$ to $R^2=1$.

We found that the biases that result from this misspecification grow slowly with
 the magnitude of the contribution of the interaction terms.
  In particular, the estimated complementarities in the~\can{Choo-Siow:06} model
    are remarkably
   robust 
  to the inclusion of interaction terms.

\subsubsection{Some Theory}\label{ssub:some_theory}
We do not know of any study of the properties of matching under non-separability, even in the limit case where $R^2=1$.  However, we state below two results and a conjecture.

Our first result builds on standard properties of linear programs in Euclidean spaces. As is well-known, the solution to any such program is generically unique and robust to small changes in the parameters. To put it more formally, define a general linear program as
\[
\max_{\bm{x}\in \bkR^n} \; \bm{c}^\prime \bm{x}  \; \mbox{ s.t. } \; \bm{A}\bm{x}\leq\bm{b}
\]
and assume that $\bm{A}$ and $\bm{b}$ define a non-empty constrained set. Then the problem has a solution set $\bm{x}(\bm{A},\bm{b},\bm{c})$. For a generic choice of $(\bm{A},\bm{b},\bm{c})$, 
the solution is a singleton; and if $(\bm{A}^\prime,\bm{b}^\prime,\bm{c}^\prime)$ is close enough to $(\bm{A},\bm{b},\bm{c})$, 
\[
\bm{x}(\bm{A}^\prime,\bm{b}^\prime,\bm{c}^\prime)=\bm{x}(\bm{A},\bm{b},\bm{c}).
\]
Whether it is separable or not, TU matching is an instance of a linear program; and it is finite-dimensional if the number of potential partners is finite.  Therefore for a generic draw of the parameters, the optimal matching is a piecewise constant function of $\sigma$. In  particular, it is generically the same for fully separable and for almost separable models:

\bigskip

\textbf{Result 1:} in finite markets, generically (for almost all draws),
 there is a $\bar{\sigma}$ such that the optimal matching is the same 
 for all $0\leq \sigma<\bar{\sigma}$.

\bigskip

Result 1 exploits the fact that in finite markets, any statistic of the model
 that is a continuous function of $\sigma$ is locally constant. This does not extend to  the ``large markets'' limit
 that is usually assumed in the empirical applications of separable matching models: then such statistics become smooth
  functions of $\sigma$. 
    Our second result states that for small $\sigma$, these statistics 
    (and in particular any misspecification bias) are of order $O(\sigma^2)$ for small $\sigma$ when
     the non-separable component is drawn from a distribution that is symmetric around zero.

\medskip

{\bf Result 2:}  in large markets, if  the distribution of 
$\nu$ is symmetric around zero then the effects of non-separability are in
$\sigma^2$ for small $\sigma$.

\medskip

The proof of result 2 relies on the fact that in large markets, the empirical distribution of 
all draws of the $\nu_{ij}$ component converges to its generating distribution. If the latter is symmetric, then 
changing all $\sigma\nu_{ij}$ to $(-\sigma)\nu_{ij}$ cannot affect the equilibrium in the large market limit. A fortiori, 
all statistics computed on the equilibrium must be locally even functions of $\sigma$. The 
equilibrium matching is a smooth function of $\sigma$ in the limit; therefore its difference 
 with the equilibrium matching of the separable model must be at most $O(\sigma^2)$.

In general, adding a non-separable mean-zero term in a separable model is akin 
to increasing the opportunities for successful matches. By making the market ``thicker'', 
it increases the probabilities of all kinds of matches: we expect (and our simulations
will confirm)  $\mu_{11}, \mu_{12}, \mu_{21},$ and $\mu_{22}$ to be larger, 
and $\mu_{10}, \mu_{01}, \mu_{20}$ and $\mu_{02}$ to be smaller, than in the fully 
separable model.

On the other hand, since we are adding random terms that are distributed 
independently of partners' observed types, there is no obvious reason why the probabilities of non-diagonal
 matches ($\mu_{12}$ and $\mu_{21}$) should increase much differently than those of diagonal 
 matches ($\mu_{11}$ and $\mu_{22}$). In particular, the log-odds-ratio
 \[
   \log\frac{\mu_{11}\mu_{22}}{\mu_{12}\mu_{21}}
   \]
   may not be so different in non-separable models. In the \can{Choo-Siow:06} specification that we
   use to estimate the parameters of the model, this ratio identifies with 
the
``supermodular core'' defined in 
\can{csw:17}, which in this simple case is simply the double difference
\[
  D_2\Phi\equiv \Phi_{11}+\Phi_{22}-\Phi_{12}-\Phi_{21}.
  \]
From this admittedly imprecise argument we derive the following conjecture:

\medskip

 {\bf Conjecture:} mistakenly assuming  separability generates 
 ``small'' misspecification biases  on  the estimated supermodular core $D_2\Phi$
 of the joint surplus.

\subsubsection{Monte Carlo Setup}\label{ssub:simulation}
To test our conjecture, we  simulated average-size samples of individuals with varying
 characteristics and matching preferences. 
Our  simulation framework takes in 1,000 individuals, equally split between men and women\footnote{We will also report
more briefly on a scaled-down population of 200 individuals.}.  We assign to each one of two educational levels; 
 we examine both symmetric and asymmetric margins. 
In the symmetric case, there are 250 men and 250 women in 
each of the two groups (HS and CG). 
The asymmetric scenario assumes a larger number of college 
educated women (375) than men (125); conversely, 
it has a smaller number of high-school educated women (125)
 than men (375).

Recall that we defined  
a ``non-separability''  $R^2$. We calibrate our scenarii so that 
this $R^2$ takes values
 of $0, 0.2, 0.4, 0.6, 0.8$,
 and $1$. This covers the range from the fully separable 
 \can{Choo-Siow:06} model  to a ``fully random'' surplus (drawn 
 independently of the observed types). 
 The   intermediate cases represent a mildly
  non-separable model ($R^2=0.2, 0.4$) and a strongly 
  non-separable one ($R^2=0.6, 0.8$).

All of our simulations pre-impose a supermodular and
 symmetric systematic surplus function $\Phi$. We distinguish two cases: 
 ``small modularity'', with
 \begin{equation*}
  \Phi = 
  \begin{pmatrix}
        0.5 & 1.0  \\
        1.0 & 1.6  \\
      \end{pmatrix}
  \end{equation*}
  and ``large modularity'', where we use
  \begin{equation*}
    \Phi = 
    \begin{pmatrix}
          0.5 & 1.0  \\
          1.0 & 2.5
        \end{pmatrix}.
    \end{equation*}
    Note that the only difference lies in the surplus generated by a 
    couple of college-educated partners, which is larger in the latter case.
    As a consequence, the supermodular core
    \[
      D_2\Phi\equiv \Phi_{11}+\Phi_{22}-\Phi_{12}-\Phi_{21}
      \]
      equals $0.1$ with small modularity and $1.0$ with large modularity.

As a result of selecting two different specifications of modularity and two contrasting population margins, 
we arrive at four distinct calibrations of the variance of the systematic surplus $\Phi$. 
In the ``large market'' limit of the separable \can{Choo-Siow:06} model when $R^{2}=0$, 
the variance of $\Phi$ in the case of small modularity is
	about $0.357$ with symmetric margins and
	roughly $0.336$ with asymmetric margins.
In the case of large modularity, they are more than twice as high:
  about $0.856$ with symmetric margins and roughly $0.716$ 
  with asymmetric margins\footnote{In this paragraph, we compute the variance as $E_{\mu} \Phi_{XY}^2-(E_{\mu} \Phi_{XY})^2$, where $\mu$ is the stable matching in the large markets limit with $R^2=0$.}.

To complete the description of our simulations, we need to describe the specification of the $\varepsilon, \eta$, and 
(for the non-fully separable cases $R^2>0$) also $\nu$.  We choose to draw all $\varepsilon^i_y, \eta^j_x$, and $\nu_{ij}$ 
independently from the centered standard type I extreme value distribution such that when $R^2=0$, this is just
 the \can{Choo-Siow:06}  model; scenarii with positive $R^2$ explore its robustness 
 to deviations from separability.

For each simulation scenario, we generate 1,000 datasets.
Table \ref{tab:simulation_parameters} summarizes the simulation
 scenarii.

\begin{table}[htb]
\centering
\scalebox{0.7}{
\begin{tabular}{lllrrrrr}
\toprule
\toprule
Population: & 1,000 & & & (Separable)      & $R^2 = 0$\\
Draws: & 1,000 & & & (Non-Separable)  & $R^2 = 0.2, 0.4, 0.6, 0.8, 1$\\
Modularity: & 
Small or Large  & & &

  \\[0.5cm]
  & &  \multicolumn{2}{c}{\bf Symmetric Margins} 
  & \multicolumn{2}{c}{\bf Asymmetric Margins} \\
\cmidrule(r){3-4}
\cmidrule(r){5-6}
                    & &               & Share of     &                & Share of\\
                    & & Count      & Population
                     &  Count     & Population \\ 
\midrule
Men & HS \hspace{0.15cm}  ($x = 1$)   & 250   & 25\%  & 375 & 37.5\%   \\
Men & CG \hspace{0.08cm} ($x = 2$)   & 250   & 25\%  & 125 & 12.5\% \\
\midrule
Women & HS \hspace{0.15cm} ($y = 1$)    & 250   & 25\%  & 125 & 12.5\%   \\
Women & CG \hspace{0.08cm} ($y = 2$)   & 250   & 25\%  & 375 & 37.5\%      \\
\bottomrule 
\end{tabular}
}
\captionof{table}{Simulation Parameters}
\label{tab:simulation_parameters}
\end{table}

\subsubsection{Monte Carlo Results}\label{ssub:results}
The combination of six values of $R^2$, 
 symmetric or asymmetric populations, and 
 two modularity subcases generates 24 different scenarii.
  As going through 
  all of our results would quickly bore the reader, we focus on the 
  most striking ones. Sometimes we only show plots for the
  small modularity/symmetric margins case to save space.
  
  With non-separable surplus, the only way we know of solving for the equilibrium matching is 
  to solve the linear programming problem associated with the primal (maximizing the total surplus) or dual
  (minimizing the sum of utilities under the stability constraints).
  We used the \texttt{R} interface of the free academic version 
  of the Gurobi software\footnote{\can{gurobi:manual}.} for this purpose. The algorithm converges very robustly.
  Even with our relatively small populations of 1,000 individuals, 
  each run requires several gigabytes of memory. This stands in contrast with the separable case, for
  which very efficient methods exist---most notably the IPFP 
  algorithm of \can{cupid:22}.

The $R^2=0$ simulation serves as our benchmark, since in 
that case the model is well-specified. Biases and 
inefficiencies introduced by the 
misspecifications for $R^2>0$ will show up in a translation and
 a spreading out 
of the estimated density of our estimators of the four elements 
of the $\Phi$ matrix, 
and of the supermodular core $D_2\Phi$.

Figures~\ref{appxGumbelSmallModAllPhiSym} and~\ref{appxGumbelLargeModAllPhiSym} 
plot the distributions of the estimated $\hat{\Phi}_{xy}$ for respectively
$(x,y)=(1,1), (1,2), (2,1),$ and $(2,2)$. 
As our discussion in section~\ref{ssub:some_theory}
suggested,  the estimators have a positive bias that grows with the extent
of the non-separability. This reflects the growing thickness of
 the market and the resulting higher probability of finding a
  suitable partner. This is apparent in
   Figure~\ref{appxGumbelSmallSymSurpluses}: as the joint surplus becomes more
    non-separable, each group gets better outcomes.

 More interestingly, Figures~\ref{GumbelSmallD2PhiSym} and~\ref{GumbelLargeD2PhiSym}
 show that our estimate of the supermodular core $D_2\Phi$ 
 has very little bias, even when $R^2$ becomes as large as $0.6$. 
  Recall that for this value of $R^2$, the non-separable term 
  $\nu_{ij}$ contributes half more variance
  than the sum of the separable terms $\varepsilon^i_y$ and $\eta^j_x$. 
  This finding is important as we have known since \can{becker:73} that 
  the supermodularity of the joint surplus drives the essential properties of the matching. 
  It points to a remarkable robustness of the \can{Choo-Siow:06} estimator
   in the face of rather large
  deviations from separability.

 Another important property of the matching 
equilibrium are its ``equilibrium prices'':  how it shares
 the  joint surplus between the two partners in any couple that
 forms in equilibrium. Figures~\ref{GumbelSmallSymShares} and~\ref{GumbelLargeSymShares}
 show the distribution
 of the share that goes to the man on average, that is
\begin{equation*}
  \frac{u_x}{u_x+v_y}
\end{equation*}
for all types of couples
 (depending whether either partner is  college-educated). We focus
 on the symmetric case, where the shares should be close to $0.5$ when
 the partners are equally educated. Both figures confirm this; again, the lack of bias
 in the estimated surplus shares is striking, even with $R^2=1$ when the 
 basic structure of the \can{Choo-Siow:06} model vanishes. 

 With symmetric populations and supermodularity, one expects the couples in which one partner is better-educated than the
 other to attribute a smaller share of the surplus to the less-educated partner; and the difference should
 grow with the supermodular core. Comparing Figures~\ref{GumbelSmallSymShares} and~\ref{GumbelLargeSymShares}
 shows that while men do almost as well with $x=1,y=2$ as with $x=2, y=1$ when the modularity is small,
 the difference becomes more noticeable with large modularity.

 Since solving the linear programming problem is much
more expensive than using the IPFP algorithm
proposed by~\can{cupid:22}, it is also interesting to compare 
their performances in our simulation runs. While we already know that
IPFP is much faster (and uses very little memory), it is only
 rigorously valid in the ``large market'' limit 
 when the number of individuals goes to infinity. Figure~\ref{GumbelSmallSymIPFPperformance}
 shows that even with our  population size of 1,000, 
 IPFP performs remarkably well: it yields equilibrium matching probabilities 
 $\mu_{xy}$ that are very close to the mode of the distribution of
 the Gurobi results for the finite-population case.

  We also ran a set of simulations with only 200 individuals. 
 The results are basically similar as with 1,000 individuals, with more variation
 across runs as expected.

\incgraff{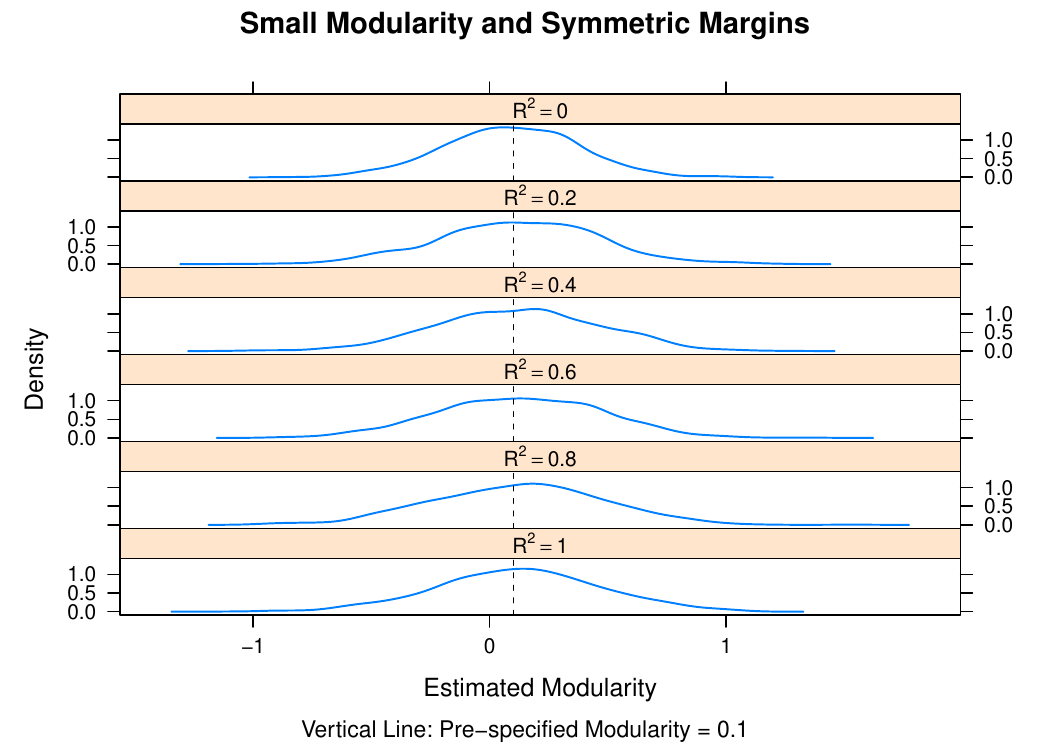}{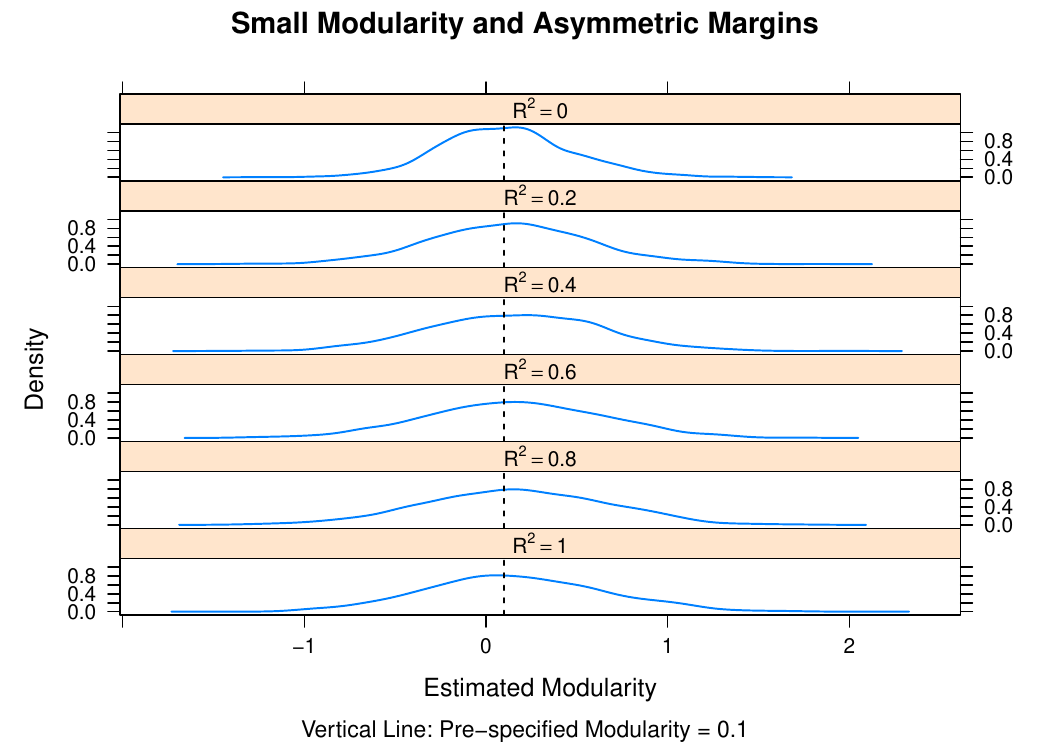}{Estimates of $D_2\Phi$---small modularity}{width=0.87\textwidth}

\incgraff{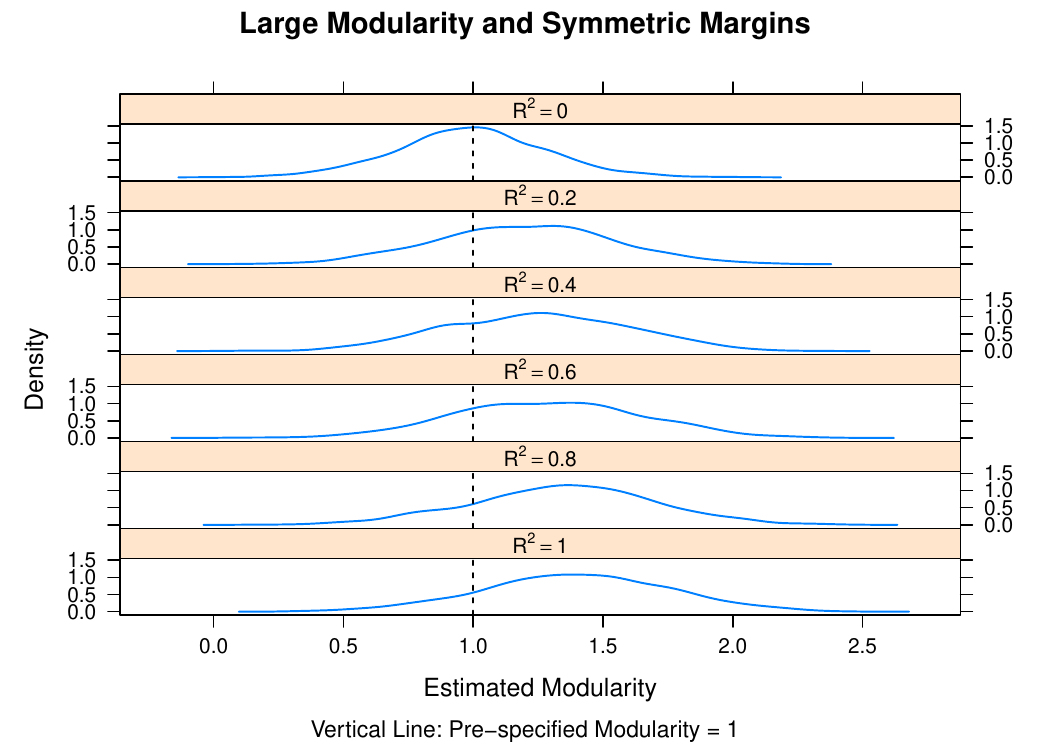}{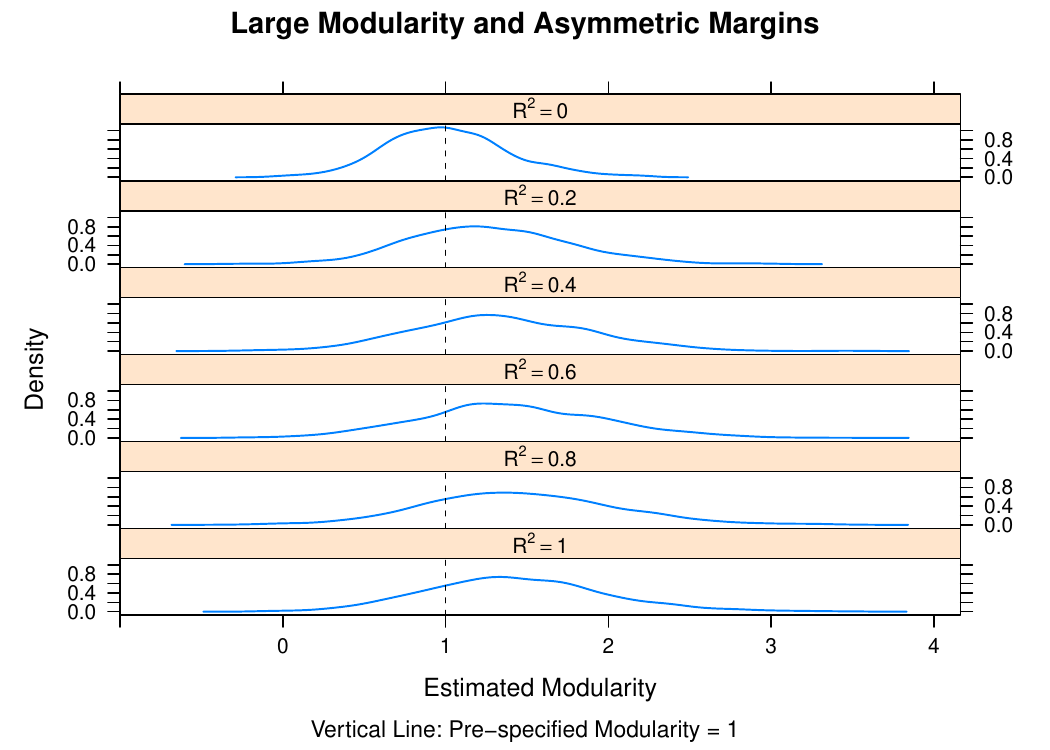}{Estimates of $D_2\Phi$---large modularity}{width=0.87\textwidth}

\incgraf{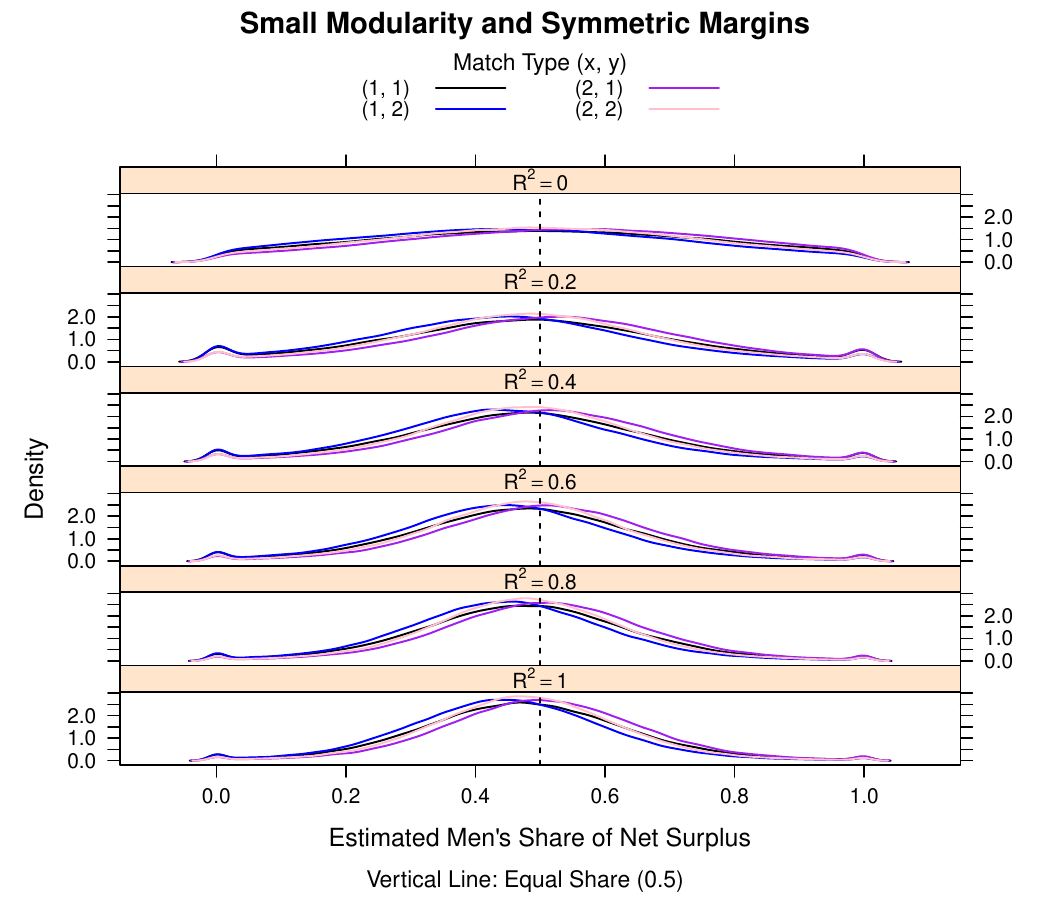}{Estimates of Men's Shares---symmetric, small modularity}{width=\textwidth}

\incgraf{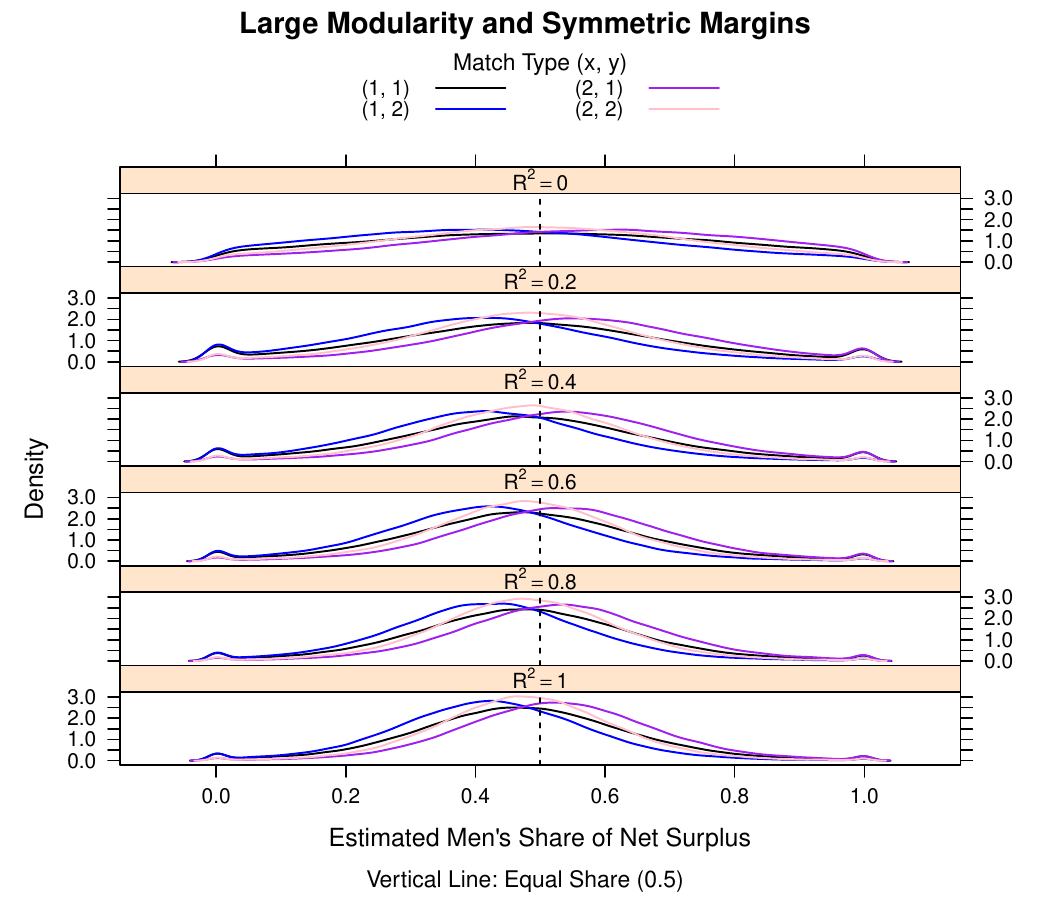}{Estimates of Men's Shares---symmetric, large modularity}{width=\textwidth}

\incgraf{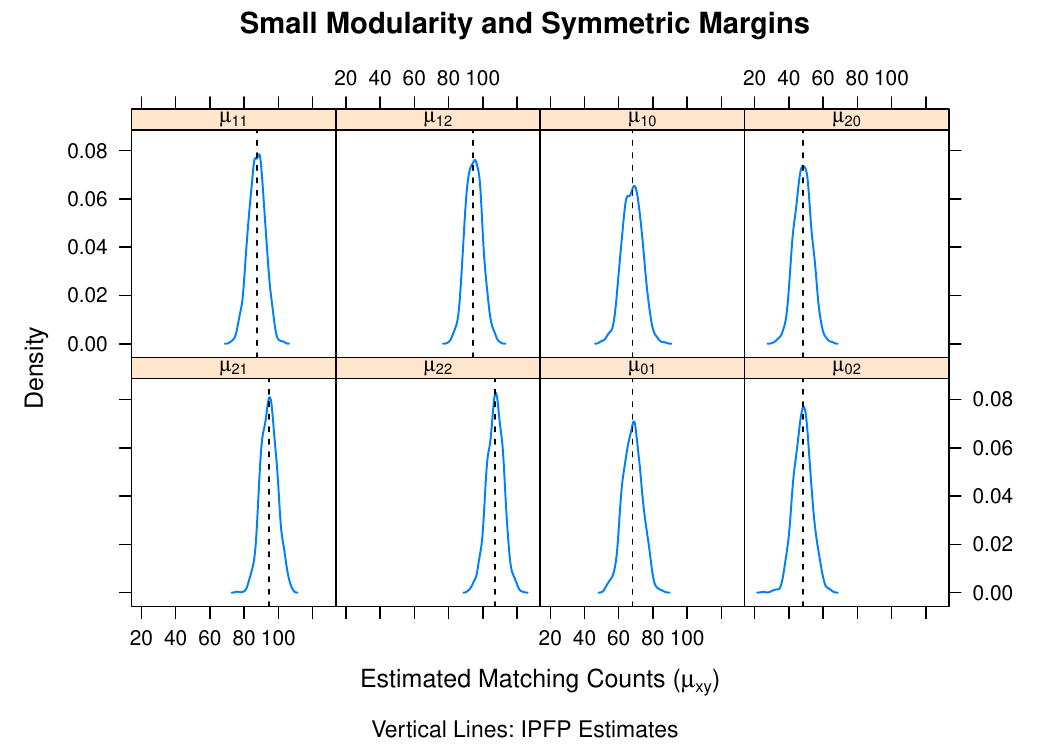}{Estimates of $\mu_{xy}$---symmetric, small modularity}{width=\textwidth}

\newpage
\section{A Caveat on Imbalanced Data}\label{sec:caveat_on_imbalanced_data}
The estimation of matching models, separable or not, is very sensitive to the presence of  large variation in cell sizes. 
To take a very bare-bones example, consider a Choo and Siow model with $X=Y=2$ observed types 
and suppose that we have no data on singles: we only observe the proportion of marriages  $0<\hat{\mu}_{xy}<1$ 
in each cell $x,y=1,2$. Then we can only estimate the double difference
\[
\phi_0 = \Phi_{11}+\Phi_{22}-\Phi_{12}-\Phi_{21}. 
\]
The maximum likelihood estimator of $\phi_0$ is simply 
\[
\hat{\phi}= 2\log\frac{\hat{\mu}_{11}\hat{\mu}_{22}}{\hat{\mu}_{12}\hat{\mu}_{21}}.
\]
It is easy to check that  given a sample of $n$ marriages, the asymptotic approximation to its variance is 
\[
V\hat{\phi}_n= \frac4n \left(\frac{1}{\mu_{11}}+\frac{1}{\mu_{12}}
+\frac{1}{\mu_{21}}+\frac{1}{\mu_{22}}\right),
\]
where the $\mu_{ij}$ are the asymptotic limits of the equilibrium proportions. 
It is clear in this form that since $\mu_{11}+\mu_{12}+\mu_{21}+\mu_{22}=1$, 
the variance of the estimator is always at least $64/n$. To get an estimate of $\phi_0$ with a standard error
of say $0.1$, we need $6,400$ marriages in the best of cases (that is, when $\phi_0=0$ so that all $\mu_{ij}$ equal $1/4$.).
It it not hard to show that this bound also holds with $X$  equally probable types on each side (rather than just two)\footnote{To be precise, the variance is $16/(nD_0(1-D_0))\geq 64/n$, where
$D_0$ is the proportion of marriages with $x=y$.}.

Even more importantly, a larger dispersion in cell sizes increases the variance of the estimator. In fact,
\[
V\hat{\phi}_n > \frac4n \frac{1}{\min(\mu_{11}, \mu_{12}, \mu_{21},\mu_{22})}:
\]
the presence of {\em any\/} small cell will make the estimator imprecise.

It may not be  so surprising, a posteriori, that estimating a multinomial
 two-sided choice model requires fairly large samples; and we know that even
  in one-sided models, small cells create problems for estimation. This difficulty is  more salient in matching models, as cells with small numbers of observations are often found in  the data. Marriage markets are a good illustration: e.g.\ \can{Choo-Siow:06} used data with individuals ages 16 to 75, with non-zero cells clustered near the diagonal. \citeasnoun[Section 5.1]{estsep:24} is a first stab at some possible strategies to deal with zero cells. Non-zero but small cells may be easier to deal with: they could be trimmed out of the data, cautiously inflated, or weighted 
down in the estimation. This is an important topic that clearly requires more work.

\section*{Concluding Remarks}
This paper provides a balanced assessment of the empirical strengths and limitations of separable matching models.
Our theoretical arguments and our Monte Carlo simulations suggest (cautiously) that omitted variables and omitted non-separabilities   induce misspecification biases that
do not seem to be severe. Estimating a separable, \can{Choo-Siow:06} model on data that 
may have been generated  by a non-separable model (or a separable model with some unobserved variables) does little apparent harm to our ability
 to get reliable  estimates of the most economically important statistics: the supermodular core
 and the surplus shares. 

Our main caveat is that when the data contain cells that vary widely in size, estimators are likely to be very imprecise. It may in fact be best to exclude some small cells from the estimation, to inflate their size, or to smooth cell sizes in some way. As such data are a common occurrence, more work is needed to explore which of these strategies are most useful.

\bibliography{Matching.bib}

\newpage

\appendix
\setcounter{figure}{0}
\renewcommand{\thefigure}{A.\arabic{figure}}

\begin{center}
{\large \bf Appendix: Additional Monte Carlo Results}
\end{center}
\vspace{-0.25cm}

\incgraffapp{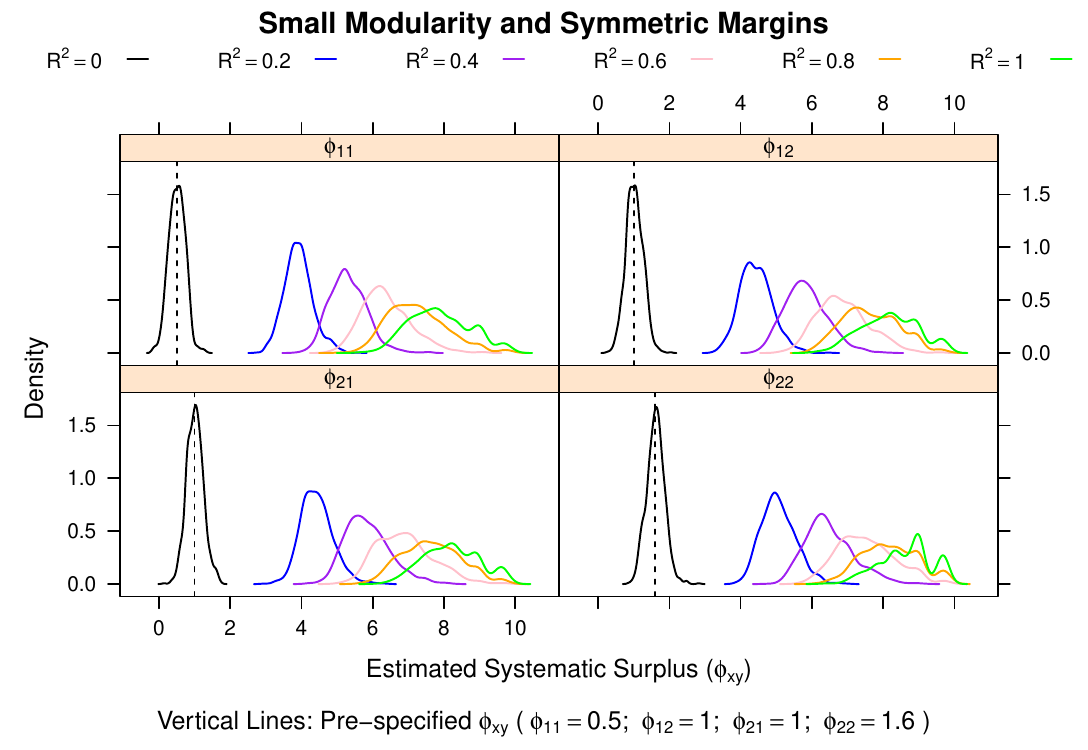}{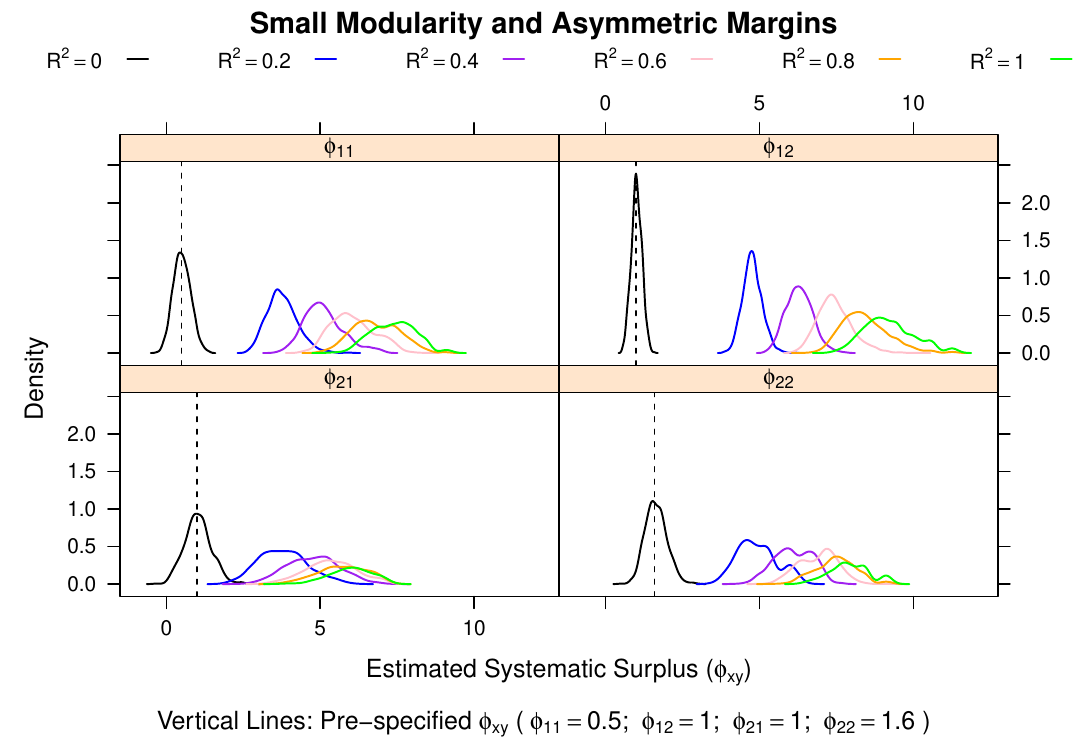}{Estimates of $\Phi$---small modularity}{width=0.825\textwidth}

\incgraffapp{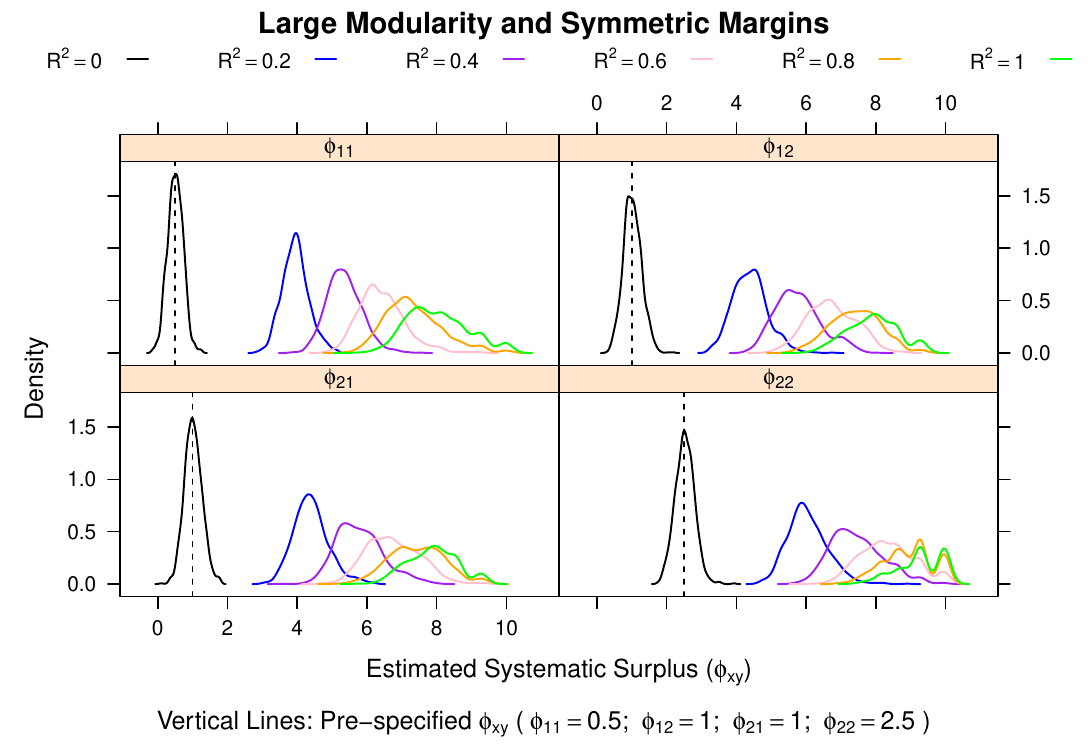}{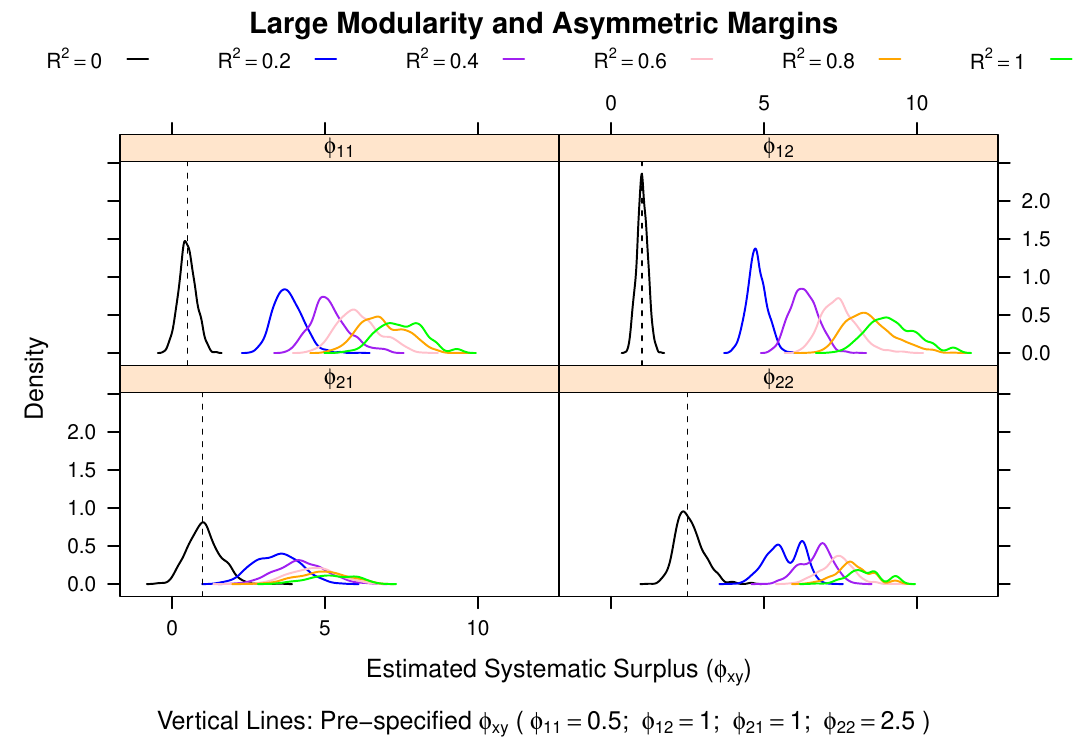}{Estimates of $\Phi$---large modularity}{width=0.825\textwidth}

\incgraf{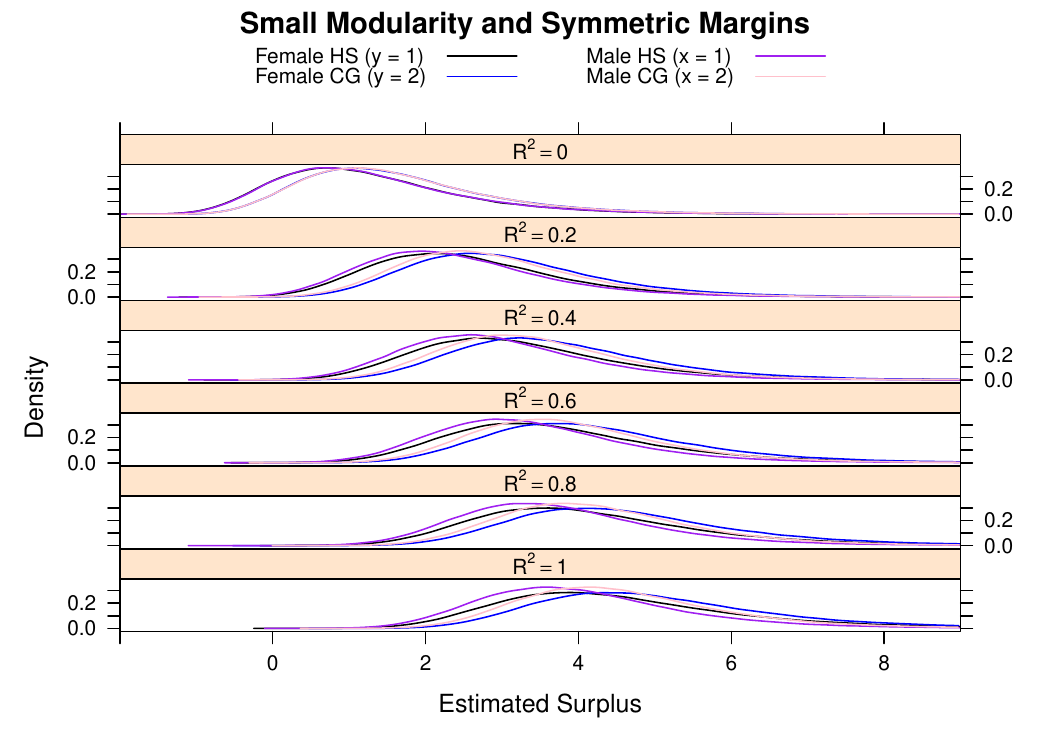}{Estimates of $u_x$ and $v_y$---symmetric, small modularity}{width=\textwidth}

\end{document}